\documentclass{llncs}

\usepackage[table]{xcolor}

\usepackage{amssymb}
\usepackage{graphicx}
\usepackage{url}
\usepackage[T1]{fontenc}
\usepackage[latin1]{inputenc}
\usepackage{lmodern}
\usepackage{paralist}
\usepackage{tikz}
\usepackage{xspace}
\usepackage[lofdepth,lotdepth]{subfig}
\usepackage[linesnumbered,lined,boxed,commentsnumbered,ruled,vlined]{algorithm2e}
\SetAlFnt{\small\sffamily}

\usepackage{graphics}
\usepackage{latexsym}
\usepackage{newlfont}
\usepackage{mathrsfs}

\usepackage{epstopdf}

\usepackage{listings}
\usepackage{array,multirow,graphicx}
\usepackage{anyfontsize}

\usepackage{caption}

\usepackage[french]{todonotes} 

\usepackage{moreverb}

\usepackage{tabularx}  
\usepackage{ragged2e}  
\newcolumntype{Y}{>{\RaggedRight\arraybackslash}X} 
\newcolumntype{Z}{>{\centering\arraybackslash}X}
\newcolumntype{L}[1]{>{\raggedright\let\newline\\\arraybackslash\hspace{0pt}}m{#1}}

\newcommand{\feta}[0]{\textsf{FETA}\xspace}
\newcommand{\lift}[0]{\textsf{LIFT}\xspace}

\def\part{\mathcal{P}}

\usepackage{tabularx}  
\usepackage{ragged2e}  

\DeclareCaptionType{copyrightbox}
\newcolumntype{Y}{>{\RaggedRight\arraybackslash}X} 
\newcommand{\tuple}[1]{\langle #1 \rangle}

\def\iri{\mathcal{I}}
\def\literal{\mathcal{L}}

\def\part{\mathcal{P}}

 % ...

\makeatletter
\providecommand*{\toclevel@algorithm}{0}
\makeatother
\def\email#1{{\tt#1}}

\usetikzlibrary{matrix,chains, positioning,shapes.geometric,arrows,backgrounds,decorations.markings,shadows,calc,shapes.misc}

 \tikzstyle{arrow} = [thick,->,>=stealth]

\newcommand{\tikzmark}[1]{\tikz[overlay,remember picture] \node (#1) {};}

\newcommand*{\DrawArrow}[3][]{%
    \begin{tikzpicture}[overlay,remember picture]
        \draw [very thick, -stealth, #1] ($(#2)$) to ($(#3)$);
    \end{tikzpicture}%
}%

\begin{document}

\title{Extracting Basic Graph Patterns from Triple Pattern Fragment Logs}

\author{Georges Nassopoulos \and Patricia Serrano-Alvarado \and Pascal Molli \and \\  Emmanuel Desmontils}
\institute{LS2N Laboratory -  Universit\'e de Nantes -- France \\ \email{\{firstname.lastname\}@univ-nantes.fr}}

\maketitle
\begin{abstract}

  The Triple Pattern Fragment (TPF) approach is de-facto
  a new way to publish Linked Data at low cost and with high server
  availability.  However, data providers hosting TPF servers are not
  able to analyze the SPARQL queries they execute because they only receive and evaluate
  queries with one triple pattern. In this paper, we propose \lift: an
  algorithm to extract Basic Graph Patterns (BGPs) of executed queries from
  TPF server logs.  
  Experiments show that \lift 
  extracts BGPs with good precision and good recall generating limited noise.

\end{abstract}


\section{Introduction}
\label{sec:introduction}

The Triple Pattern Fragment (TPF) approach is de-facto a new
way to publish Linked Data at low cost and with high availability for
data
providers~\cite{verborgh_jws_2016}. WarDrobe~\cite{wardrobe_beek_14}
provides more than 38 billions of triples distributed over 65
data\-sets. Following the TPF approach, most of the SPARQL query
processing is now executed on the client-side, \emph{TPF servers only
receive and evaluate queries with one triple pattern}.  Consequently, data providers of 
TPF servers do not know the executed SPARQL queries and cannot
analyze them as data providers do with queries of SPARQL endpoints.

Knowing executed SPARQL queries is fundamental for
data providers. Mining logs of SPARQL endpoints allows to
detect recurrent patterns in queries for
prefetching~\cite{detecting_sparql_templates_lorey_13} or for
benchmarking~\cite{morsey2011dbpedia}. It provides the type of queries
issued, the complexity and the used
resources/predicates~\cite{moller2010learning,picalausa2011real}. It
allows also to distinguish between man or machine made queries
\cite{raghuveer2012characterizing,rietveld2014man}. Currently, such
analysis cannot be done on logs of TPF servers because they only contain
information about single triple patterns.

In this paper, we propose \lift (LInked data Fragment Tracking): an algorithm to extract Basic Graph
Patterns (BGPs) from logs of TPF servers. Compared to the state of
art, \cite{verborgh_usewod_2015} reported statistics from the logs of the DBpedia's TPF server. 
In previous work~\cite{nassopoulos2016feta}, we proposed an algorithm to extract
BGPs of \emph{federated SPARQL queries} from logs of a \emph{federation of
SPARQL endpoints}. Here, we address a similar scientific problem but in
the context of a single TPF server.

The main challenge to extract BGPs is the concurrent
execution of SPARQL queries on one TPF server. If we find a function
$f$, to extract BGPs from isolated traces of one SPARQL query, is $f$
able to extract the same BGP from traces of concurrent SPARQL queries?
\lift faces this problem by tracking the bindings among different
triple pattern queries to detect joins. We experimented \lift with different levels of
concurrency. We demonstrate in which conditions, it 
extracts BGPs with good precision and good recall generating limited noise.  Thanks to \lift, we
were able to extract the frequent BGPs from the TPF log published in the USEWOD
2016 dataset \cite{usewod16}.

Next section
introduces a motivating example and our
problem statement. Section~\ref{sec:feta} presents LIFT
 and Section~\ref{sec:experiments} shows our experiments.
Section
\ref{sec:related_work} presents related work. Finally, conclusions and
future work are outlined in Section~\ref{sec:conclusion}.

\section{Motivating example and problem statement}
\label{sec:motivation_problem}

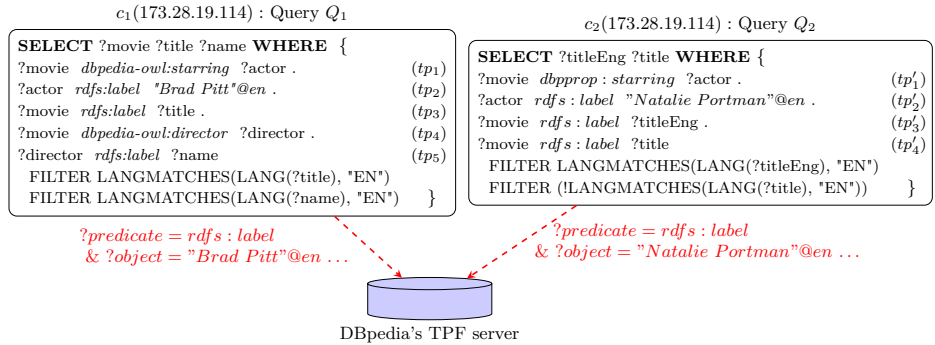
\begin{figure*}
   \centering \scalebox{.75}{
\begin{tikzpicture}
\centering

\tikzstyle{vecArrow} = [thick, decoration={markings,mark=at position
   1 with {\arrow[semithick]{open triangle 60}}},
   double distance=3pt, shorten >= 5.5pt,
   preaction = {decorate},
   postaction = {draw,line width=2pt, white,shorten >= 4.5pt}]

\node(q1)at(-3,0)[draw, align=left, rounded corners, text width=7.7cm, 
font=\small, label=above:$c_1 (173.28.19.114):$ Query $Q_1$]{
\begin{tabular}{lc}
  {\fontsize{8}{12}\selectfont  \textbf{SELECT} ?movie ?title ?name \textbf{WHERE} } \{  & \\
  {\fontsize{8}{12}\selectfont  ?movie ~\emph{dbpedia-owl:starring} ~?actor  . } &   {\fontsize{8}{12}\selectfont  ($tp_1$) } \\
  {\fontsize{8}{12}\selectfont  ?actor ~\emph{rdfs:label} ~\emph{"Brad~Pitt"@en} . } &   {\fontsize{8}{12}\selectfont  ($tp_2$) } \\
  {\fontsize{8}{12}\selectfont  ?movie ~\emph{rdfs:label} ~?title . } &   {\fontsize{8}{12}\selectfont ($tp_3$) } \\
  {\fontsize{8}{12}\selectfont  ?movie ~\emph{dbpedia-owl:director} ~?director . } &   {\fontsize{8}{12}\selectfont  ($tp_4$) } \\
  {\fontsize{8}{12}\selectfont  ?director ~\emph{rdfs:label} ~?name  } &   {\fontsize{8}{12}\selectfont  ($tp_5$) } \\

         {\fontsize{8}{12}\selectfont  ~ FILTER LANGMATCHES(LANG(?title), "EN") } & \\ 
        {\fontsize{8}{12}\selectfont  ~ FILTER LANGMATCHES(LANG(?name),  "EN")  } & \}
  \end{tabular}};

\node(q2)at(5.3,0)[draw, align=left, rounded corners, text width=8cm, 
font=\small,label=above:$c_2 (173.28.19.114):$ Query $Q_2$]{
\begin{tabular}{lc}
       {\fontsize{8}{12}\selectfont \textbf{SELECT} ?titleEng ?title \textbf{WHERE}} \{  & \\ 
        {\fontsize{8}{12}\selectfont  ?movie ~$dbpprop:starring$  ~?actor .} &   {\fontsize{8}{12}\selectfont  ($tp'_1$) } \\
        {\fontsize{8}{12}\selectfont  ?actor ~$rdfs:label$ ~$"Natalie~Portman"@en$ . } &   {\fontsize{8}{12}\selectfont  ($tp'_2$) } \\
      {\fontsize{8}{12}\selectfont  ?movie ~$rdfs:label$ ~?titleEng . } &   {\fontsize{8}{12}\selectfont  ($tp'_3$) } \\
       {\fontsize{8}{12}\selectfont ?movie ~$rdfs:label$ ~?title  } &   {\fontsize{8}{12}\selectfont  ($tp'_4$) } \\ 
        {\fontsize{8}{12}\selectfont   ~ FILTER LANGMATCHES(LANG(?titleEng), "EN") } & \\ 
         {\fontsize{8}{12}\selectfont  ~ FILTER (!LANGMATCHES(LANG(?title),  "EN")) } & \}
  \end{tabular}};


\node (DBpedia) [cylinder,
                cylinder uses custom fill,
                cylinder body fill=blue!20,
                cylinder end fill=blue!20, 
                shape border rotate=90,
                draw,
                 minimum height=0.75cm,
                 minimum width=2.2cm, 
                 label={[align=center]below:DBpedia's TPF server}] at (0.5,-3.2) {};

 \path[arrow](q1) edge[red,dashed] node[left,  text width=5cm]{
   $?predicate=rdfs:label$\\ $~\&~?object="Brad~Pitt"@en$ $\ldots$}  (DBpedia);

 \path[arrow](q2) edge[red,dashed] node[right, text width=6cm]{
   $~~~~?predicate=rdfs:label$\\ $~\&~?object="Natalie ~Portman"@en$ $\ldots$ } (DBpedia);

\end{tikzpicture}}
   \caption{Concurrent execution of queries $Q_1$ and $Q_2$.}
   \label{fig:motiv1}
 \end{figure*}

In Figure~\ref{fig:motiv1}, two clients, $c_1$ and $c_2$, execute
concurrently queries $Q_1$ and $Q_2$ over the DBpedia's TPF
server.  $Q_1$ asks for movies starring
Brad Pitt and $Q_2$ for movies starring Natalie Portman.\footnote{These 
queries come from \url{http://client.linkeddatafragments.org/}.}
Both queries have one BGP composed of several triple patterns ($tp_{n}$). 

 TPF clients decompose SPARQL queries into a sequence of triple
 pattern queries.
 Table~\ref{tab:query_logs_dbpedia_q1} presents some traces of
 the TPF server for query
 $Q_1$. Odd-numbered lines represent received triple pattern queries and 
 even-numbered ones represent sent triples after evaluation on the RDF graph.

\begin{table*}
 \centering
\begin{footnotesize}
\begin{tabularx}{\textwidth}{|l| @{} | l | p{1.2cm}  | Y | p{1cm} @{}}  \hline \rowcolor{gray!70}
 & \textbf{IP}  & \textbf{Time} & \textbf{Asked triple pattern/TPF}  \\ \hline

1& 172...& 11:24:19  &  ?predicate=rdfs:label~\&~?object="Brad Pitt"@en \\\rowcolor{gray!40}
2& 172...& 11:24:23  &  \textbf{dbpedia:Brad\_Pitt}\tikzmark{mapping1} rdfs:label "Brad Pitt"@en ,  \\\hline  
3& 172...& 11:24:24  & ?predicate=dbpedia-owl:starring~\&~?object=\textbf{dbpedia:Brad\_Pitt}\tikzmark{injected1}  \\\rowcolor{gray!40}
4& 172... & 11:24:27  &   \textbf{dbpedia:A\_River\_Runs\_Through\_It\_(film)}\tikzmark{mapping2} dbpedia-owl:starring  dbpedia:Brad\_Pitt \\  \rowcolor{gray!40}
&& & dbpedia:Troy\_(film) dbpedia-owl:starring dbpedia:Brad\_Pitt  ...\\  \hline
5&172...& 11:24:28  & ?subject=\textbf{dbpedia:A\_River\_Runs\_Through\_It\_(film)}\tikzmark{injected2-1} \&?predicate=rdfs:label \\ \hline 
 \end{tabularx}
\end{footnotesize}
   \caption{Excerpt of log of the DBpedia's TPF server for query $Q_1$.}
   \label{tab:query_logs_dbpedia_q1}
\end{table*}


 

Lines 1 and 3, correspond to triple pattern queries for $tp_2$ and $tp_1$ of  $Q_1$.\footnote{TPF clients only 
request bound parts of a triple patterns, variables are omitted.} 
We can observe that the object in Line 3, 
comes from a mapping seen in Line 2. 
This \emph{injection} of a mapping obtained from a previous triple pattern query, is 
clearly a join implemented in a nested-loop from $tp_2$ towards $tp_1$. 
 


As the TPF server only sees triple pattern queries, the original
queries e.g., $Q_1$ and $Q_2$ are unknown to the data provider. In
this work, we address the following research question: \emph{Can we
  extract the BGPs from a TPF server log?} 
  
  In our example, we aim to extract two BGPs from the TPF server log, one corresponding to $Q_1$,
BGP[1]$=\{ tp_1.tp_2.tp_3.\-tp_4.tp_5\}$ and another corresponding to
$Q_2$, BGP[2]$=\{tp'_1.tp'_2.tp'_3.tp'_4\}$.
Before presenting our scientific problem,  we introduce the following definitions.

\begin{definition}[BGP] \label{eq:bgp} A BGP (Basic~Graph~Pattern) is a set of triple patterns. Any tuple $\in ( \iri \cup \literal \cup V) \times (\iri \cup V) \times (\iri \cup \literal \cup V)$ is a triple pattern $\tuple{s,p,o}$, where $\iri$ is the set of all IRIs, $\literal$ the set of all literals and $V$ the set of all variables disjoint from $\literal$ and $\iri$.\footnote{We do not consider blank nodes in this paper.} 
\end{definition}


\begin{definition}[TPF server log] \label{eq:ldflog} A TPF server log is a totally ordered sequence of execution traces structured in tuples $\tuple{ip,ts,tp,\mu_o}$ where \emph{ip} is the IP address of the client, \emph{ts} is the timestamp of the http request, \emph{tp} is a triple pattern, and $\mu_o$ is the set of RDF triples returned by the TPF server transformed in mappings.
\end{definition}

We denote by $E(Q_i)$, the log produced by a TPF server when evaluating the SPARQL query $Q_i$ and by 
$E(Q_1 \parallel ...  \parallel Q_n)$ the log of \emph{n concurrent queries}.


%
\begin{definition}[Approximation $\approx$ of BGPs]
A BGP approximates another  ($\approx$) if, to some extent, both contain same triple patterns 
and same joins. 
\end{definition}

To measure such approximation we can use \emph{precision} and  \emph{recall} of triple patterns and joins 
of one BGP against another. The average of precision and recall can be used as a measure of global \emph{quality} of the approximation.

\begin{definition}[Problem of BGP reversing] \label{eq:1} Given a log corresponding to the execution of one query, $E(Q_i)$, find a function $f(E(Q_i))$ producing a set of BGPs $\{BGP_1,...,\-BGP_n\}$,  such that:
\begin{itemize}
	\item [Property 1.] $f(E(Q_i))$ approximates ($\approx$) the BGPs existing in the original query. If we consider that $BGP(Q_i)$ returns the set of BGPs of $Q_i$ then $f(E(Q_i)) \approx BGP(Q_i)$.
	\item [Property 2.] $f(E(Q_i))$ guarantees \emph{resistance to concurrency}, i.e., BGPs obtained from the log of \emph{isolated} queries, approximate ($\approx$) results obtained from the log of \emph{concurrent} queries:
$f(E(Q_1))\cup ... \cup f(E(Q_n)) \approx f(E(Q_1\parallel...\parallel Q_n)).$
\end{itemize}
\end{definition}

{
\def\OldPoint{.}
\catcode`\.=13
\def.{%
  \ifmmode%
  \OldPoint\discretionary{}{}{}%
  \else%
  \OldPoint%
  \fi%
}%

We evaluate the BGPs extracted by $f$ with the precision, recall and quality of triple patterns and joins 
returned by $f$ against  those existing in original queries. 
If $f(E(Q_1))$ extracts the BGP$=\{tp_1.tp_2.tp_3.tp_4.tp_5\}$, then precision, recall and quality of triple patterns and joins are 
perfect according to the BGP present in $Q_1$,  even if variables have different names. 
But if $f(E(Q_1))$ misses one triple pattern (e.g., $tp_5$), then precision is $4/4$, recall is $4/5$ and quality is $(1+4/5)/2$.

}



{ \def\OldPoint{.}  \catcode`\.=13 \def.{%
    \ifmmode%
    \OldPoint\discretionary{}{}{}%
    \else%
    \OldPoint%
    \fi%
  }%
  In Figure~\ref{fig:motiv1}, if $c_1$ and $c_2$ have different IP
  addresses, it is possible to separate $E(Q_1 \parallel Q_2)$ into
  $E(Q_1)$, $E(Q_2)$ and apply the reversing function to each
  trace. However, in the worst case, $c_1$ and $c_2$ have the same IP
  address, i.e., a web application running on the cloud that executes 
  queries $Q_1$ and $Q_2$ in parallel. Thus, we expect that
  $f(E(Q_1 \parallel Q_2)) \approx f(E(Q_1)) \cup f(E(Q_2))$.  }

\section{LIFT: a reversing function}
\label{sec:feta}

We propose \lift as a system of heuristics to implement $f$. The idea
is to detect nested-loop joins. In Table
\ref{tab:query_logs_dbpedia_q1}, the mappings returned in Line 2 are
reused in the next triple pattern query at Line 3. We track such
bindings in order to link variables of different triple pattern
queries.
In this paper, we make the following hypothesis:
\begin{inparaenum}[(i)]
\item we consider only bound predicates,
\item we do not consider the server's web cache (this information 
   can be easily obtained by data providers), and
\item we do not consider the client's TPF cache. 
\end{inparaenum}
The first hypothesis can be omitted, but we kept it because analysis of SPARQL 
queries show that predicates are frequently bound \cite{gallego2011empirical}. 

\begin{footnotesize}
\begin{figure}
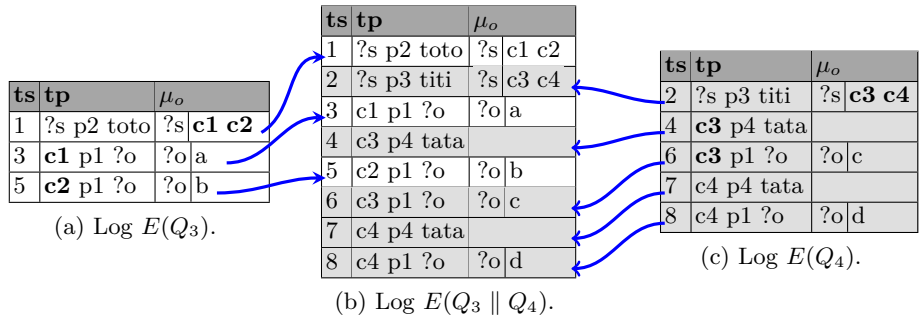

  \centering

\subfloat[Log $E(Q_3)$.]{\label{fig:logQ3} 
\begin{tabular}{|l|l|l|} \hline \rowcolor{gray!70}

\textbf{ts} &\textbf{tp} & \textbf{$\mu_o$}  \\ \hline
1  & ?s~p2~toto &
\begin{tabular}{l|ll}
?s & \textbf{c1} & \textbf{c2} \tikzmark{SQ31}\\
\end{tabular}
\\ \hline
3 & \textbf{c1}~p1~?o &
\begin{tabular}{l|ll}
?o & a &  \\ 
\end{tabular} \tikzmark{SQ32}\\ \hline 
5 & \textbf{c2}~p1~?o &
\begin{tabular}{l|l}
?o & b \tikzmark{SQ33}\\
\end{tabular} \\ \hline
\end{tabular}}\hfill%
\subfloat[Log $E(Q_3 \parallel Q_4)$.]{\label{fig:extractlogmix} 
\begin{tabular}{|l|l|ll|} \hline \rowcolor{gray!70}

\textbf{ts} &\textbf{tp} & \textbf{$\mu_o$} & \\ \hline

\tikzmark{log1}1 & ?s~p2~toto &
\begin{tabular}{l|ll}
?s & c1 & c2 \\
\end{tabular}
& \\ \hline \rowcolor{gray!25}

2 & ?s~p3~titi &
\begin{tabular}{l|ll}
?s & c3 & c4  
\end{tabular}
& \tikzmark{log2} \\ \hline

\tikzmark{log3}3 & c1~p1~?o &
\begin{tabular}{l|ll}
?o & a &  \\ 
\end{tabular} & \\ \hline \rowcolor{gray!25}

4 & c3~p4~tata & 
& \tikzmark{log4} \\ \hline

\tikzmark{log5}5 & c2~p1~?o &
\begin{tabular}{l|l}
?o & b\\ 
\end{tabular}
& \\ \hline \rowcolor{gray!25}

6 & c3~p1~?o &
\begin{tabular}{l|l}
?o & c 
\end{tabular}
& \tikzmark{log6} \\ \hline \rowcolor{gray!25}

7 & c4~p4~tata &
\begin{tabular}{l|l}
\end{tabular} 
& \tikzmark{log7} \\ \hline \rowcolor{gray!25}

8 & c4~p1~?o & 
\begin{tabular}{l|l}
?o & d 
\end{tabular} 
& \tikzmark{log8} \\\hline

\end{tabular}}\hfill%
\DrawArrow[blue, out=30, in=180]{SQ31}{log1} 
\DrawArrow[blue, out=0, in=180]{SQ32}{log3}
\DrawArrow[blue, out=0, in=180]{SQ33}{log5}  
\subfloat[Log $E(Q_4)$.]{\label{fig:logQ4} 
\begingroup\setlength{\fboxsep}{0pt}
\colorbox{gray!25}{%
\begin{tabular}{|l|l|l|} \hline \rowcolor{gray!70}

\textbf{ts} &\textbf{tp} & \textbf{$\mu_o$} \\ \hline
\tikzmark{SQ41}2 & ?s~p3~titi &
\begin{tabular}{l|ll}
?s & \textbf{c3} & \textbf{c4} \\ 
\end{tabular}
\\ \hline
\tikzmark{SQ42}4 & \textbf{c3}~p4~tata & \\ \hline
\tikzmark{SQ43}6 & \textbf{c3}~p1~?o &
\begin{tabular}{l|l}
?o & c \\ 
\end{tabular}
\\ \hline
\tikzmark{SQ44}7 & c4~p4~tata &
\begin{tabular}{l|l}
\end{tabular}
\\ \hline 
\tikzmark{SQ45}8 & c4~p1~?o & 
\begin{tabular}{l|l}
?o & d \\ 
\end{tabular} \\\hline
\end{tabular}}%
\endgroup
}

\DrawArrow[<-,blue, out=0, in=180]{log2}{SQ41}  
\DrawArrow[<-,blue, out=0, in=180]{log4}{SQ42}  
\DrawArrow[<-,blue, out=0, in=180]{log6}{SQ43}  
\DrawArrow[<-,blue, out=0, in=180]{log7}{SQ44}  
\DrawArrow[<-,blue, out=0, in=180]{log8}{SQ45}  

\caption{Examples of simplified TPF logs.}
\label{fig:ex2-log}
\end{figure}
\end{footnotesize}


Figure~\ref{fig:ex2-log} presents a simplified log of $E(Q_{3})$, $E(Q_{4})$ and $E(Q_3 \parallel Q_4)$ where:
\begin{inparaenum}
\item[]  \\$Q_3=SELECT\ *\ WHERE \ \{ ?x \ p2 \ toto \ . \  ?x \ p1 \ ?y \}$ and
\item[]  \\$Q_4=SELECT\ *\ WHERE \ \{?x \ p3\ titi\ . \ ?x \ p1\ ?y \ . \ ?x \ p4 \ tata \}$.
\end{inparaenum}

For the sake of simplicity, timestamps are transformed into
integers. The IP address of the TPF client is the same for $Q_3$ and $Q_4$, so we
removed the $ip$ column.  Variables are named $?s$ or
$?o$. $\mu_o$ represents the mappings of variables resulting from the
evaluation of $tp$. We call them \emph{output-mappings}.
 Observe the client first requests more selective triple patterns, i.e., $\tuple{?x \ p2 \ toto}
 $ for $Q_{3}$ and $\tuple{?x \ p3 \ titi}$ for $Q_{4}$, leaving at the end less selective 
 ones i.e., $\tuple{?x \ p1 \ ?y}$ for both queries. 
 Then mappings returned by selective patterns are 
 bound into less selective ones producing a nested-loop. See that mappings $c1$ and 
 $c2$ are bound in the variable $?x$ of the second triple pattern of $Q_{3}$. Similarly, 
 mappings $c3$ and $c4$ are bound in the variable $?x$ of the other triple 
 patterns of $Q_4$. We call \emph{input-mappings} these injected mappings. 
 Modified triple patterns are the inner part of the nested-loop that
 we call \emph{inner loop}. We call  \emph{outer loop} the triple patterns whose mappings 
 are used to bound variables, e.g., $\tuple{?x \ p2 \ toto}$.
 

The basic intuition of \lift is to detect if mappings obtained in a request are bound in
next requests. This can be challenging because mappings can be :
\begin{inparaenum}[(i)]
\item bound several times (e.g., in star queries), 
\item bound partially as a side-effect of LIMIT and FILTER
  clauses,
\item or bound into a different concurrent query.
\end{inparaenum}

As a real log can be huge, \lift analyzes the log on a sliding window
defined by a \emph{gap}, i.e., a time interval. When \lift reads an
entry $e$ in the log with a timestamp $ts$, it considers only entries
reachable within the gap i.e., $ts \pm gap$. 
Algorithm~\ref{global_algo} shows the three phases of \lift.
\vspace{-0.3cm}
\IncMargin{1em}
\SetInd{0.25em}{0.25em}
\begin{algorithm}
\DontPrintSemicolon
\SetKwInOut{Input}{input}\SetKwInOut{Output}{output}\SetKwInOut{Data}{data}
\SetKwProg{Fn}{Function}{ is}{}

\Fn{LIFT($log, gap$)}{
\Input{a TPF server log; a $gap$ in time units (seconds)}
\Output{a set of $BGPs$}
\Data{$CTP$ a list of ctps, $DTP$ a graph of dtps}
\SetKw{CTPExtraction}{ctpExtraction}
\SetKw{nlDetection}{nestedLoopDetection}
\SetKw{bgpC}{bgpExtraction}

\BlankLine
 $CTP \gets$ \CTPExtraction($log, gap$)\;
 $DTP \gets$ \nlDetection($CTP, gap$)\;
 return $BGP \gets \bgpC(DTP)$
 }
\caption{Global algorithm of LIFT}
\label{global_algo}
\end{algorithm}\DecMargin{1em}

\vspace{-0.3cm}
\begin{enumerate}
\item First, \lift \textbf{merges} triple pattern queries having same characteristics into \emph{candidate triple
  patterns (ctp)}. This allows to gather triple pattern queries that
  seem to be part of the same inner loop.
\item Next, \lift looks for an inclusion relationship among
  output-mappings and input-mappings of ctps. If such an inclusion exists a  \emph{deduced triple
  pattern (dtp)} is created. If instead of inclusion an intersection exists, \lift
  \textbf{splits} ctps to obtain a dtp with inclusion. 
  If neither inclusion nor intersection exists an isolated dtp is created. This produces a 
   $DTP$ Graph where nodes are dtps and edges are inclusion relationships between dtps.
\item Finally, \lift extracts BGPs from the DTP graph.
  Ideally, $\lift(E(Q_3 \parallel Q_4), gap)$ should compute the 2 BPGs of $Q_3$ and $Q_4$:\\
  $\{?s\ p2\ toto\ \ . \ ?s \ p1 \ ?o\},\{?s\ p3\ titi \ . \ ?s \ p1 \
  ?o \ .\ ?s \ p4 \ tata\}$.
\end{enumerate}
Section~\ref{sec:extr-cand-triple} details the CTP extraction. Section
\ref{sec:nest-loop-detect} describes the nested-loop detection.
Finally, Section~\ref{sec:bgp-extraction} presents the phase of
extraction of BGPs.

\subsection{Extraction of candidate triple patterns}
\label{sec:extr-cand-triple}

The objective of $ctpExtraction$ is to aggregate together log entries
that seem to participate in the same outer or inner loop.  Aggregated entries
are ctps. 
A ctp is a tuple $\tuple{ip, ts, tp, \mu_o, \mu_i}$ where $ip$ is an
IP address, $ts$ is a pair of timestamps $(s.min, ts.max)$ representing
a range; when creating a ctp both timestamps are identical and
correspond to the timestamp of the corresponding entry in the
log. $tp$ is a triple pattern query, $\mu_o$ (output-mappings) is the
list of solution mappings for variables of
$tp$. 
$\mu_i$ (input-mappings) is a set of mappings built during the
\textit{ctpExtraction}. Basically, we replace any constant of $tp$ by
a variable, we use $\sigma$ for
subject and $\omega$ for object. 
Replaced constants are regrouped in $\mu_{i}$.
%
%
\vspace{-0.3cm}
\IncMargin{1em}
\SetInd{0.25em}{0.25em}
\begin{algorithm}
\DontPrintSemicolon
\SetKwInOut{Input}{input}\SetKwInOut{Output}{output}\SetKwInOut{Data}{data}
\SetKwProg{Fn}{Function}{ is}{}
\SetKw{ctp}{ctp}
\SetKw{simi}{similar}
\SetKw{merge}{merge}
\SetKw{ctpExtraction}{ctpExtraction}

\Fn{\ctpExtraction($log,gap$)}{
  \Input{a TPF server $log$; a $gap$ in time units (seconds)}
  \Output{a list $CTP$ of ctp}
  \BlankLine
  $CTP\gets$[ ]\;
  \ForEach{$e \in log$}{
    c $\gets$ read(e) as (ip,(ts,ts), tp, $\mu_o$, $\mu_i$)
    \Switch {c.tp}{
       \lCase{?s p o:} {c.tp $\gets$ ?s p $?o_{in}$ ; c.$\mu_i$ $\gets$ $?\omega$|o}
       \lCase{s p ?o:} {c.tp $\gets$ $?s_{in}$ p ?o ; c.$\mu_i$ $\gets$ $?\sigma$|s}
       \lCase{s p o:}  {c.tp $\gets$ $?s_{in}$ p $?o_{in}$  ; c.$\mu_i$ $\gets$ $?\sigma$|s, $?\omega$|o}
      }

      \If {$\exists$ $c_k$ $\in$ $CTP$ | ingap(c,$c_k$,gap) $\wedge$ $c_k.ip=c.ip$ $\wedge$ $c.tp=c_k.tp$}{
        $c_k$. $\mu_o$ $\cup$ c. $\mu_o$ ; $c_k$. $\mu_i$ $\cup$ c. $\mu_i$ ;
        $c_k$.ts.max=c.ts.max;
      } \lElse {CTP.add(c)}}
  return $CTP$
}
\caption{Extraction of Candidate Triple Patterns}
\label{algo:ctpExtraction}
\end{algorithm}\DecMargin{1em}
%
\vspace{-0.3cm}

\begin{small}
\begin{figure}
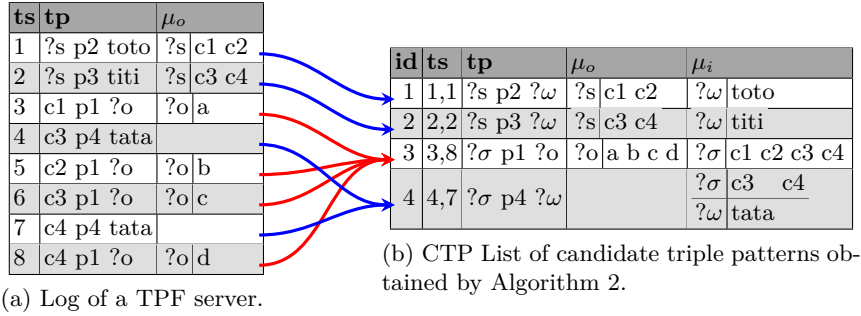

  \centering
\subfloat[Log of a TPF server.]{\label{fig:extractlog} 
\begin{tabular}{|l|l|ll|} \hline \rowcolor{gray!70}
\textbf{ts} &\textbf{tp} & \textbf{$\mu_o$} & \\ \hline
1 & ?s~p2~toto &
\begin{tabular}{l|ll}
?s & c1 & c2 \\
\end{tabular}
&\tikzmark{SQ1} \\ \hline \rowcolor{gray!25}
2 & ?s~p3~titi &
\begin{tabular}{l|ll}
?s & c3 & c4 \\ 
\end{tabular}
&\tikzmark{SQ2}\\ \hline
3 & c1~p1~?o &
\begin{tabular}{l|ll}
?o & a &  \\ 
\end{tabular} 
& \tikzmark{SQ3}\\ \hline \rowcolor{gray!25}
4 & c3~p4~tata & 
& \tikzmark{SQ4} \\ \hline
5 & c2~p1~?o &
\begin{tabular}{l|l}
?o & b\\ 
\end{tabular}
& \tikzmark{SQ5}\\ \hline \rowcolor{gray!25}
6 & c3~p1~?o &
\begin{tabular}{l|l}
?o & c \\ 
\end{tabular}
&\tikzmark{SQ6}\\ \hline
7 & c4~p4~tata &
\begin{tabular}{l|l}
\end{tabular}
&\tikzmark{SQ7} \\ \hline \rowcolor{gray!25}
8 & c4~p1~?o & 
\begin{tabular}{l|l}
?o & d \\ 
\end{tabular} 
&\tikzmark{SQ8} \\ \hline
\end{tabular}}\hfill%
\subfloat[CTP List of candidate triple patterns obtained by Algorithm \ref{algo:ctpExtraction}.]{\label{fig:extractctp}
\begin{tabular}{|l|l|l|l|l|} \hline \rowcolor{gray!70}
\textbf{id} & \textbf{ts} &\textbf{tp} & \textbf{$\mu_o$} &\textbf{$\mu_i$} \\ \hline
\tikzmark{ctp1} 1 & 1,1 & ?s~p2~$?\omega$   
& \begin{tabular}{l|lll}
  ?s & c1 & c2 &  \\ 
  \end{tabular}  
  & \begin{tabular}{l|l}
  $?\omega$ & toto \\ 
  \end{tabular} \\ \hline  \rowcolor{gray!25}

\tikzmark{ctp2} 2 & 2,2 &?s~p3~$?\omega$    
& \begin{tabular}{l|ll}
  ?s & c3 & c4 \\ 
  \end{tabular}
& \begin{tabular}{l|l}
  $?\omega$ & titi \\ 
  \end{tabular} \\ \hline

\tikzmark{ctp3} 3 & 3,8 & $?\sigma$~p1~?o   
& \begin{tabular}{l|llll}
  ?o & a & b & c & d\\
  \end{tabular}
& \begin{tabular}{l|llll}
  $?\sigma$ & c1 & c2 & c3 & c4\\ 
  \end{tabular} \\ \hline \rowcolor{gray!25}
 \rowcolor{gray!25}

\tikzmark{ctp4} 4 & 4,7 & $?\sigma$~p4~$?\omega$ 
& 
& \begin{tabular}{l|ll}
  $?\sigma$ & c3 & c4 \\ \hline
  $?\omega$ & tata & 
  \end{tabular} \\ \hline 

\end{tabular}
}
\DrawArrow[blue,out=0, in=180]{SQ1}{ctp1}
\DrawArrow[blue, out=0, in=180]{SQ2}{ctp2}
\DrawArrow[red, out=0, in=180]{SQ3}{ctp3}
\DrawArrow[red, out=0, in=180]{SQ5}{ctp3}
\DrawArrow[red, out=0, in=180]{SQ6}{ctp3}
\DrawArrow[red, out=0, in=180]{SQ8}{ctp3}
\DrawArrow[blue, out=0, in=180]{SQ4}{ctp4}
\DrawArrow[blue, out=0, in=180]{SQ7}{ctp4}

 \caption{TPF log and CTP List produced by 
  Algorithm~\ref{algo:ctpExtraction} with $E(Q_3 \parallel Q_4)$ and
  $gap=8$.}
  
\label{fig:ctpextraction}
\end{figure}
\end{small}

Algorithm \ref{algo:ctpExtraction} outlines the extraction of a CTP
List from a TPF log with a particular
$gap$. Figure~\ref{fig:ctpextraction} illustrates the effect of
executing Algorithm \ref{algo:ctpExtraction} on log
$E(Q_3 \parallel Q_4)$ with gap=8.
 The log is processed in sequential order.  Lines 5 to 7
initialize input-mappings by replacing constants by variables $\sigma$
or $\omega$.
Next, Lines 9 to 10 merge (i.e., aggregate) current ctp
with an existing and compatible one. An existing ctp is compatible if
it has the same $tp$, it is produced by the same $ip$ address, and
fits in the gap. The $ingap(c,c_k,gap)$ function returns true if
$c.ts.min - c_k.ts.max \leq gap$. If the current ctp is compatible with
an existing one, output/input-mappings and timestamps are merged. When
updating timestamps, the lower timestamp remains always the same, only
the upper timestamp can grow.  A variable of $tp$ cannot belong to
$\mu_o$ and $\mu_i$ simultaneously.


This algorithm can aggregate triple patterns that do not belong to the
same nested-loop as it is the case in our example of Figure~\ref{fig:ctpextraction}, where CTP[3]
aggregates triple patterns of $Q_{3}$ and $Q_{4}$. We suppose that
this case is not likely, especially when the gap is small but if it is
the case, next algorithm splits ctps to separate these nested-loops.

\subsection{Nested-loop join detection}
\label{sec:nest-loop-detect}

Algorithm~\ref{algo:nestedLoopDetection} describes how to link
variables of different ctps produced by
Algorithm~\ref{algo:ctpExtraction}. It builds a DTP Graph of \emph{deduced
triples patterns (dtp)} where nodes have the same structure as ctps and edges
represent a relation of inclusion between input-mappings ($\mu_{i}$)
and output-mappings ($\mu_{o}$) of 2 different dtps. 
Figure~\ref{fig:nested-dtp} presents the DTP Graph 
 produced by 
Algorithm~\ref{algo:nestedLoopDetection} with the CTP List of
Figure~\ref{fig:nested-ctp}. Dashed links represent linked variables
deduced by Algorithm~\ref{algo:nestedLoopDetection}. 
\begin{footnotesize}
\begin{figure}
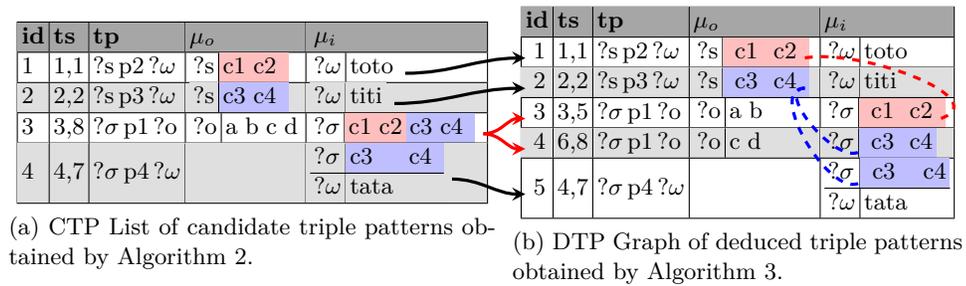

  \centering
\subfloat[CTP List of candidate triple patterns obtained by Algorithm \ref{algo:ctpExtraction}.]{\label{fig:nested-ctp}
\begin{tabular}{|l|l|l|l|l|} \hline \rowcolor{gray!70}
\textbf{id} & \textbf{ts} &\textbf{tp} & \textbf{$\mu_o$} &\textbf{\begin{tabular}{cc} $\mu_i$ \end{tabular}} \\ \hline
 1 & 1,1 & ?s\,p2\,$?\omega$   
& \begin{tabular}{l|lll}
  ?s & \cellcolor{red!25}c1 &\cellcolor{red!25}c2  \tikzmark{out1}&  \\ 
  \end{tabular}  
  & \begin{tabular}{l|l}
  $?\omega$ & toto \\ 
  \end{tabular} \tikzmark{ctp1} \\ \hline  \rowcolor{gray!25}

 2 & 2,2 &?s\,p3\,$?\omega$    
& \begin{tabular}{l|ll}
  ?s &\cellcolor{blue!25}c3 &\cellcolor{blue!25}c4  \tikzmark{out2} \\ 
  \end{tabular}
& \begin{tabular}{l|l}
  $?\omega$ & titi \\ 
  \end{tabular} \tikzmark{ctp2}\\ \hline

 3 & 3,8 & $?\sigma$\,p1\,?o   
& \begin{tabular}{l|llll}
  ?o & a & b & c & d\\
  \end{tabular}
& \begin{tabular}{l|llll}
  $?\sigma$ &\cellcolor{red!25}c1 &\cellcolor{red!25}c2 &\cellcolor{blue!25}c3 &\cellcolor{blue!25}c4 \tikzmark{in1}\\ 
  \end{tabular} \tikzmark{ctp3} \\ \hline \rowcolor{gray!25}
 \rowcolor{gray!25}

 4 & 4,7 & $?\sigma$\,p4\,$?\omega$ & 
& \begin{tabular}{l|ll}
  $?\sigma$ &\cellcolor{blue!25}c3 &\cellcolor{blue!25} c4  \tikzmark{in2}\\ \hline
  $?\omega$ & tata & 
  \end{tabular} 
\tikzmark{ctp4} \\ \hline 
\end{tabular}
}~~
\subfloat[\small{DTP Graph of deduced triple patterns obtained by Algorithm \ref{algo:nestedLoopDetection}.}]{\label{fig:nested-dtp}
\begin{tabular}{|l|l|l|l|l|} \hline \rowcolor{gray!70}

\textbf{id} & \textbf{ts} &\textbf{tp} & \textbf{$\mu_o$} &\textbf{$\mu_i$} \\ \hline

\tikzmark{dtp1} 1 & 1,1 & ?s\,p2\,$?\omega$   
& \begin{tabular}{l|lll}
 ?s &\cellcolor{red!25} c1 &\cellcolor{red!25} c2 \tikzmark{muo1}&  \\ 
  \end{tabular}  
  & \begin{tabular}{l|l}
  $?\omega$ & toto \\ 
  \end{tabular} 
\\ \hline  \rowcolor{gray!25}

\tikzmark{dtp2} 2 & 2,2 &?s\,p3\,$?\omega$    
& \begin{tabular}{l|ll}
  ?s &\cellcolor{blue!25} c3 &\cellcolor{blue!25} c4 \tikzmark{muo2} \\ 
  \end{tabular}
& \begin{tabular}{l|l}
  $?\omega$ & titi \\ 
  \end{tabular} 
\\ \hline

\tikzmark{dtp3} 3 & 3,5 & $?\sigma$\,p1\,?o   
& \begin{tabular}{l|ll}
  ?o & a & b \\
  \end{tabular}
& \begin{tabular}{l|ll}
  $?\sigma$ &\cellcolor{red!25} c1 &\cellcolor{red!25} c2 \tikzmark{mui3}\\ 
  \end{tabular} 
\\ \hline \rowcolor{gray!25}

\tikzmark{dtp4} 4 & 6,8 & $?\sigma$\,p1\,?o   
& \begin{tabular}{l|ll}
  ?o & c & d\\
  \end{tabular}
&\begin{tabular}{l|ll}
  $?\sigma$ & \tikzmark{mui4}\cellcolor{blue!25} c3 &\cellcolor{blue!25} c4\\ 
  \end{tabular} 
\\ \hline 

\tikzmark{dtp5} 5 & 4,7 & $?\sigma$\,p4\,$?\omega$ & 
& \begin{tabular}{l|ll}
  $?\sigma $ &\tikzmark{mui5}\cellcolor{blue!25} c3 &\cellcolor{blue!25} c4\\ \hline
  $?\omega $ & tata & 
  \end{tabular} 
\\ \hline 
\end{tabular}
}
\DrawArrow[black, out=0, in=180]{ctp1}{dtp1}
\DrawArrow[black, out=0, in=180]{ctp2}{dtp2}
\DrawArrow[red, out=0, in=180]{ctp3}{dtp3}
\DrawArrow[red, out=0, in=180]{ctp3}{dtp4}
\DrawArrow[black, out=0, in=180]{ctp4}{dtp5}
\DrawArrow[.-.,red,dashed, out=10, in=20]{muo1}{mui3}
\DrawArrow[.-.,blue,dashed, out=180, in=230]{muo2}{mui4} 
\DrawArrow[.-.,blue,dashed, out=180, in=230]{muo2}{mui5}

\caption{CTP List and DTP Graph
 produced by 
  Algorithm~\ref{algo:nestedLoopDetection} with $gap=8$.}
  
\label{fig:nested-ctp-dtp}
\end{figure}
\end{footnotesize}

\vspace{-0.3cm}
\IncMargin{1em}
\SetInd{0.25em}{0.25em}
\begin{algorithm}[h]
\DontPrintSemicolon
\SetKwInOut{Input}{input}\SetKwInOut{Output}{output}\SetKwInOut{Data}{data}
\SetKwProg{Fn}{Function}{ is}{}
\SetKw{nlDetection}{nestedLoopDetection}

\Fn{\nlDetection($CTP, gap$)}{
  \Input{a list $CTP$ of ctps; a $gap$ in time units (seconds)}
  \Output{An edge-labelled DTP Graph of dtps}
  
  \BlankLine
  
  \ForEach {c $\in$ CTP} {
    \lIf {split(c) $\neq$ $\emptyset$} {
      CTP.insertAfter(c.id,split(c));
    } \lElse {
    	DTP.addnode(c);
      	\ForEach {$v_o \ \in$ vars(c.$\mu_o$)} {         
        
        \ForEach {$(c_k,v_i)$ $\in$ \{ ($c_k$, $v_i$) | $c_k$ $\in$ CTP  $\wedge$ $c_k$.id $>$ c.id $\wedge$ ingap($c_k$,c,gap) $\wedge$ $\exists$ $v_i$ $\in$ vars($c_k.\mu_i$) | $c_k.\mu_i(v_i)$ $\cap$ $c.\mu_o(v_o)$ $\neq$ $\emptyset$
        \}} {
          \If { $c_k.\mu_i(v_i)$ $\subseteq$  $c.\mu_o(v_o)$ } {
            DTP.addnode($c_k$) ; DTP.addEdge(c,$c_k$,($v_o$,$v_i$));
          } \lElse { DTP.addnode(s=split($c_k,v_i,c,v_o$)) ; DTP.addEdge(c,s,($v_o$,$v_i$));
          }
        }
      } 
    }}
      return DTP;
}

\caption{Detection of nested-loop joins}
\label{algo:nestedLoopDetection}
\end{algorithm}\DecMargin{1em}
\vspace{-0.3cm}


If the $\mu_{i}$ of a ctp is a subset of the $\mu_{o}$ of a
previous ctp, then we consider that the 2 corresponding variables can be 
linked. This happens in the example described in
Figure~\ref{fig:nested-ctp} with CTP[2] and CTP[4]. We consider that
$?\sigma$ of CTP[4] is linked to ?s of CTP[2]. We formalize this
behavior at Lines 6 to 7 of Algorithm~\ref{algo:nestedLoopDetection}.

A direct inclusion does not occur if
Algorithm~\ref{algo:ctpExtraction} aggregated too many log entries as
it is the case with CTP[3]. $Q_3$ and $Q_4$ have a common
triple pattern $\tuple{?x\,p1\,?y}$ and Algorithm~\ref{algo:ctpExtraction}
aggregates them. We solve this problem by splitting a ctp.
The idea is to produce a dtp from a ctp
$c_k$ if it exists an intersection between the $\mu_{o}$ of a ctp
$c_l$ and the $\mu_{i}$ of $c_k$. In the example described in
Figure~\ref{fig:nested-ctp}, CTP[3] produces two splits: one when analyzing
CTP[1] (DTP[3] is produced) and another when analyzing CTP[2] 
(DTP[4] is produced) because both $\mu_{o}$ intersect the $\mu_{i}$ of CTP[3].  
Splitting affects input-mappings, 
but also timestamps and output-mappings. After
splitting, we obtain input-mappings that are subsets of previous
output-mappings. Intersection and splitting is shown in Lines 5 and 8 of
Algorithm~\ref{algo:nestedLoopDetection}. 

For lack of space, we do
not detail the function split. 
It basically groups, from the TPF log, values that
belong to the intersection, this generates correct timestamps,
output-mappings and input-mappings. We register the split relationship
with a $split$ predicate that links a ctp with its produced dtps. In our
example, we have 2 $split$ relations; $split(3,3)$ and
$split(3,4)$. 
Splitting has an effect on CTP traversal that we see in Line
3. Output-mappings of produced dtps must be analyzed so
when the nested-loop detection analyzes a splitted ctp, it inserts in the
CTP List, produced dtps. $split(c)$ returns the set of
dtps produced by splitting $c$.

\vspace{-0.3cm}
\subsection{BGP Extraction}
\label{sec:bgp-extraction}





\begin{figure*}
	\centering
	\includegraphics[scale=0.25] {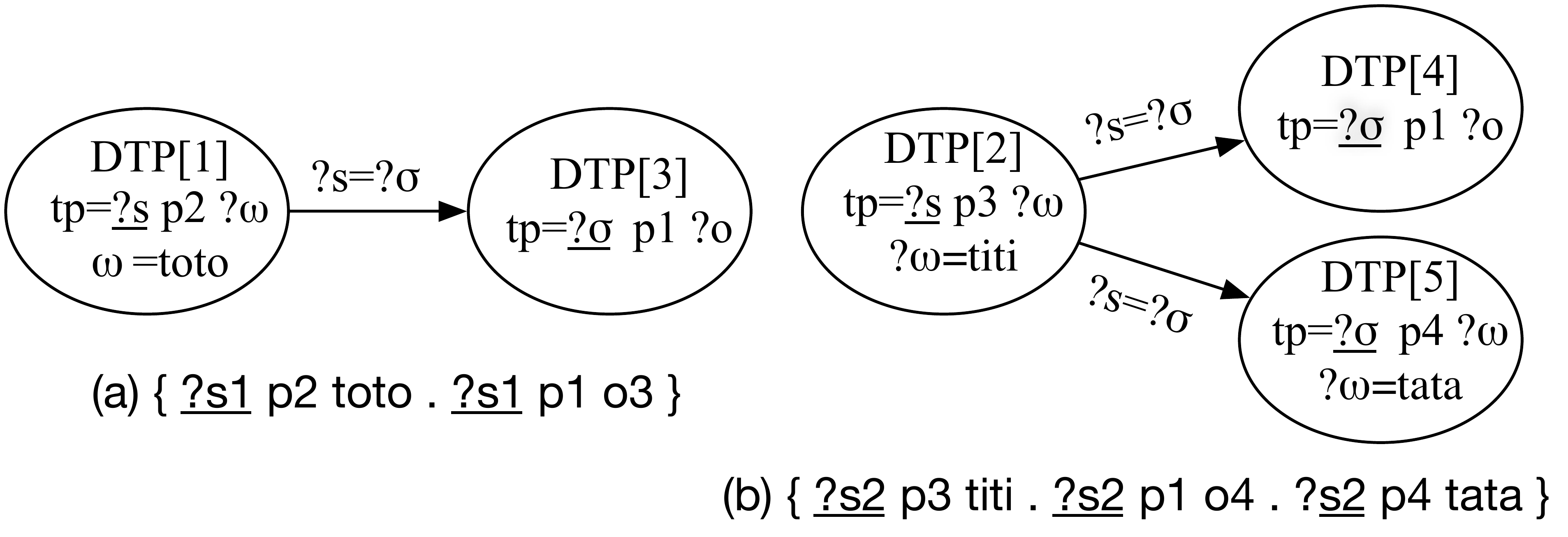}
\caption{Connected components of the DTP Graph produced by the execution of 
  Algorithm~\ref{algo:nestedLoopDetection} for $gap=8$.}
\label{fig:bgp-extract}
\end{figure*}

Figure~\ref{fig:bgp-extract} represents the connected components of
the DTP Graph shown in Figure~\ref{fig:nested-dtp}. From this
representation, it is easy to compute the final BGPs with a variable renaming
 and restitution of an IRI/literal in place of $\omega$ when there is only one
input mapping, e.g., $toto,titi$ and $tata$. Detected joins are underlined. 

In our example, \lift rebuilds perfectly BGPs of queries $Q_3$ and
$Q_4$. This example is executed with $gap=8$. If we reduce the gap,
then some joins are not detected and recall decreases. If we
execute concurrently more queries having same triple patterns, then
\lift can deduce joins that do no exist in original queries and
consequently precision will decrease. In Section~\ref{sec:experiments},
we measure experimentally the precision and recall of \lift in
different situations.

 



\section{Experiments}
\label{sec:experiments}

The goals of the experiments are twofold, 
\begin{inparaenum}[(i)]
 \item to evaluate precision and recall of \lift's results and 
 \item to show that \lift extracts meaningful BGPs from a real TPF log.  
\end{inparaenum}

First, we analyze a small set of queries taken from the TPF web site.
We evaluate precision and recall of \lift deductions
from logs of queries executed \emph{in isolation} in Section \ref{sec:isolation},
and from logs of queries executed \emph{concurrently} in Section \ref{sec:concurrent}.
These two sections evaluate to which extent \lift guarantees Properties 1 and 2 of 
our problem statement (cf. Definition \ref{eq:1}), correspondingly.

Then, we analyze \lift with an important number of real user queries taken from logs of USEWOD 2016 \cite{usewod16}.
In Section \ref{sec:endpointQueries}, we evaluate precision and recall of 14,259 queries 
coming from logs of the DBpedia's SPARQL endpoint. 
In Section \ref{sec:usewod}, we analyze 4,720,874 single triple pattern queries 
coming from logs of the DBpedia's TPF server.

\subsection{Evaluation of \lift with queries in isolation}
\label{sec:isolation}

\begin{figure}
  \centering
    \includegraphics[width=1\textwidth]{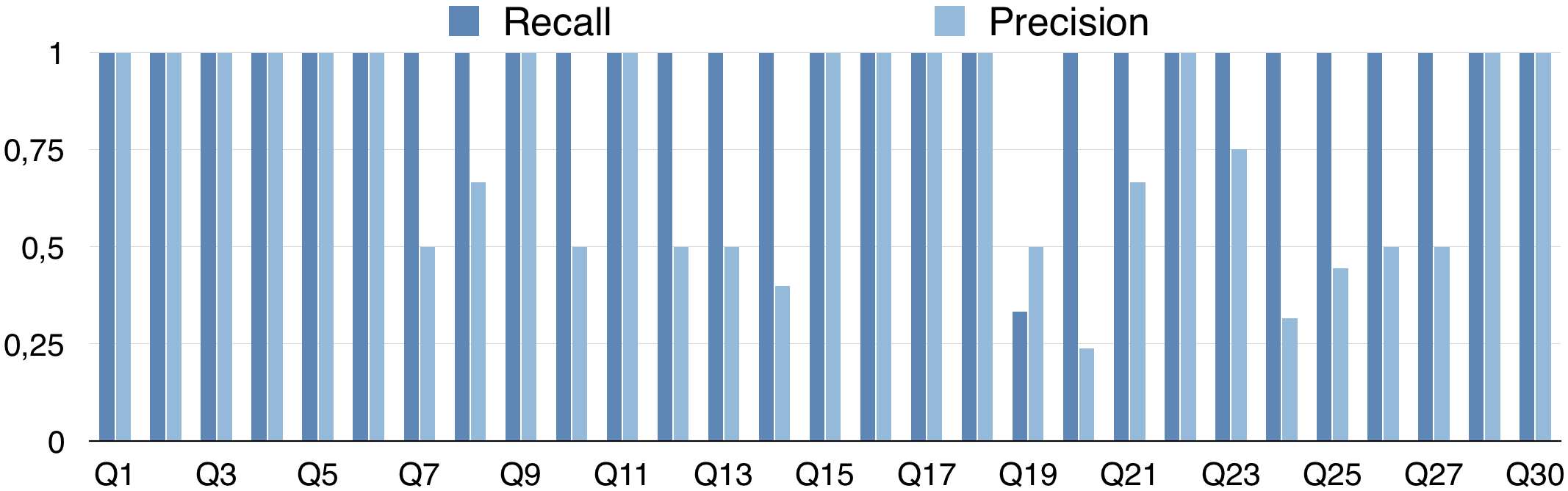}
    \caption{Precision and recall by query of the TPF web site.}
    \label{fig:isolated}
\end{figure}

We used 29 over 30 queries of the TPF web
site\footnote{\url{http://client.linkeddatafragments.org/}}.
Concerned datasets are DBpedia 2015-04, UGhent, LOV or VIAF.  We captured http
requests and answers of queries using the \emph{webInspector 1.2}
tool\footnote{\url{https://
    sourceforge.net/p/webinspector/wiki/Home/}}. 
 Source code of \lift is available on GitHub\footnote{\url{https://github.com/coumbaya/lift}}
For each query $Q_i$, we run $\lift(E(Q_i),
\infty)$.
 Figure~\ref{fig:isolated} presents precision and recall
of \lift deductions against original queries\footnote{Queries, TPF logs and \lift results are
  available at: \\ \url{https://github.com/coumbaya/lift/blob/master/experiments.md}},
  they show to which extent $\lift(E(Q_i)) \approx BGP(Q_i)$ (cf. Definition \ref{eq:1}, Property 1). 
   In average, \lift obtained 97\% of recall and 75\%
of precision of joins.  That gives a quality of 84.5\%. \lift deduces perfectly 15 of 30
BGPs: $Q_{1}-Q_{6}$, $Q_{9}$, $Q_{11}$, $Q_{15}- Q_{18}$, $Q_{22}$,
and $Q_{29}-Q_{30}$. $Q_{28}$ was not analyzed because it needs two TPF servers.


\vspace{-0.3cm}
\lift does not detect UNION queries because they are processed on the client side.
$Q_9$ is a query with a BGP like \{(tp1 UNION tp2) . tp3\}. In this case, \lift
detects 2 BGPs, \{tp1 . tp3\} and \{tp2 . tp3\}. $Q_{29}$ is also a
UNION query but without join, as there is no intersection between
mappings, \lift detects two separate BGPs containing one triple pattern. We consider this
behaviour
correct.  


Figure~\ref{fig:q7} describes $Q_{7}$ and its deduced BGPs.  BGP[1] is
correct, while BGP[2] is not. 
This is due to traces of the first request made by TPF clients for each triple
pattern of each query to decide the join ordering  
(TPF servers send cardinalities of concerned triples in their answers). 
Thus, when processing $Q_7$, the client makes a first request
 of each triple pattern and then decides to begin with
the first triple pattern. Then it binds resulting mappings into the
$?book$ variable of the second triple pattern to retrieve
corresponding authors.  This nested-loop is deduced in BGP[1].  But as
output mappings of the first request of the second triple pattern intersects
with the values of the inner loop, \lift deduces BGP[2] with a
self-join that is very unlikely and that can be easily filtered in a
post-processing.  Such situation appears in 6 of 29 queries:
$Q_{7}$, $Q_{12-14}$, $Q_{21}$ and $Q_{25}$.

 \begin{small}
\begin{figure}
  \centering
    \begin{tabular}{|l|l|l|} \hline
ID & Original query & Deduced BGPs \\ \hline
$Q_7$  &
\begin{lstlisting}[
    basicstyle=\footnotesize, %\tiny or or \small or \footnotesize etc.
]
SELECT DISTINCT 
  ?book ?author 
WHERE {
 ?book rdf:type dbpo:Book . 
 ?book dbpo:author ?author  
} LIMIT 100
\end{lstlisting}  

  &
\begin{lstlisting}[
    basicstyle=\footnotesize, %\tiny or or \small or \footnotesize etc.
]
BGP[1]:
 {?s1 rdf:type  dbpo:Book .
 ?s1 dbpo:author ?o2}
BGP[2]: 
 {?s3 dbpo:author ?o3 .
 ?s3 dbpo:author ?o4}
\end{lstlisting}   \\ \hline

$Q_{8}$ &
\begin{lstlisting}[
    basicstyle=\footnotesize, %\tiny or or \small or \footnotesize etc.
]
{SELECT ?award WHERE {
 ?award a dbpedia-owl:Award . 
 ?award dbpprop:country ?language . 
 ?language dbpedia-owl:language 
          dbpedia:Dutch_language}
\end{lstlisting}   
&
\begin{lstlisting}[
    basicstyle=\footnotesize, %\tiny or or \small or \footnotesize etc.
]
{?s1 dbpedia-owl:language 
     dbpedia:Dutch_language .
 ?s2 dbpprop:country ?s1 .
 ?s2 rdf:type dbpedia-owl:Award .
 ?s1 rdf:type dbpedia-owl:Award}
 \end{lstlisting}   \\ \hline

\end{tabular}

\caption{\lift deductions for  $Q_7$ and $Q_{8}$. Prefix dbpo corresponds to dbpedia-owl.}
\label{fig:q7}
\end{figure}
\end{small}

In some cases, \lift deduces additional triple patterns and thus false joins
with well deduced triple patterns because of the intersection of mappings. 
This is more challenging to filter. In Figure \ref{fig:q7}, $Q_{8}$ has an
additional triple pattern, the last one, and a join with the second triple pattern. 
This is the case for $Q_{8}$, $Q_{10}$, $Q_{14}$, $Q_{20}$, 
$Q_{23-27}$. 

In addition, \lift merges triple patterns when they are very similar
as it is the case in $Q_{19}$ and $Q_{20}$ where some triple patterns
have same predicate and variables in the same position
(subject/object).

\subsection{Does LIFT results resist to concurrency?}
\label{sec:concurrent}

\begin{figure}
  \centering
     \includegraphics[width=1\textwidth]{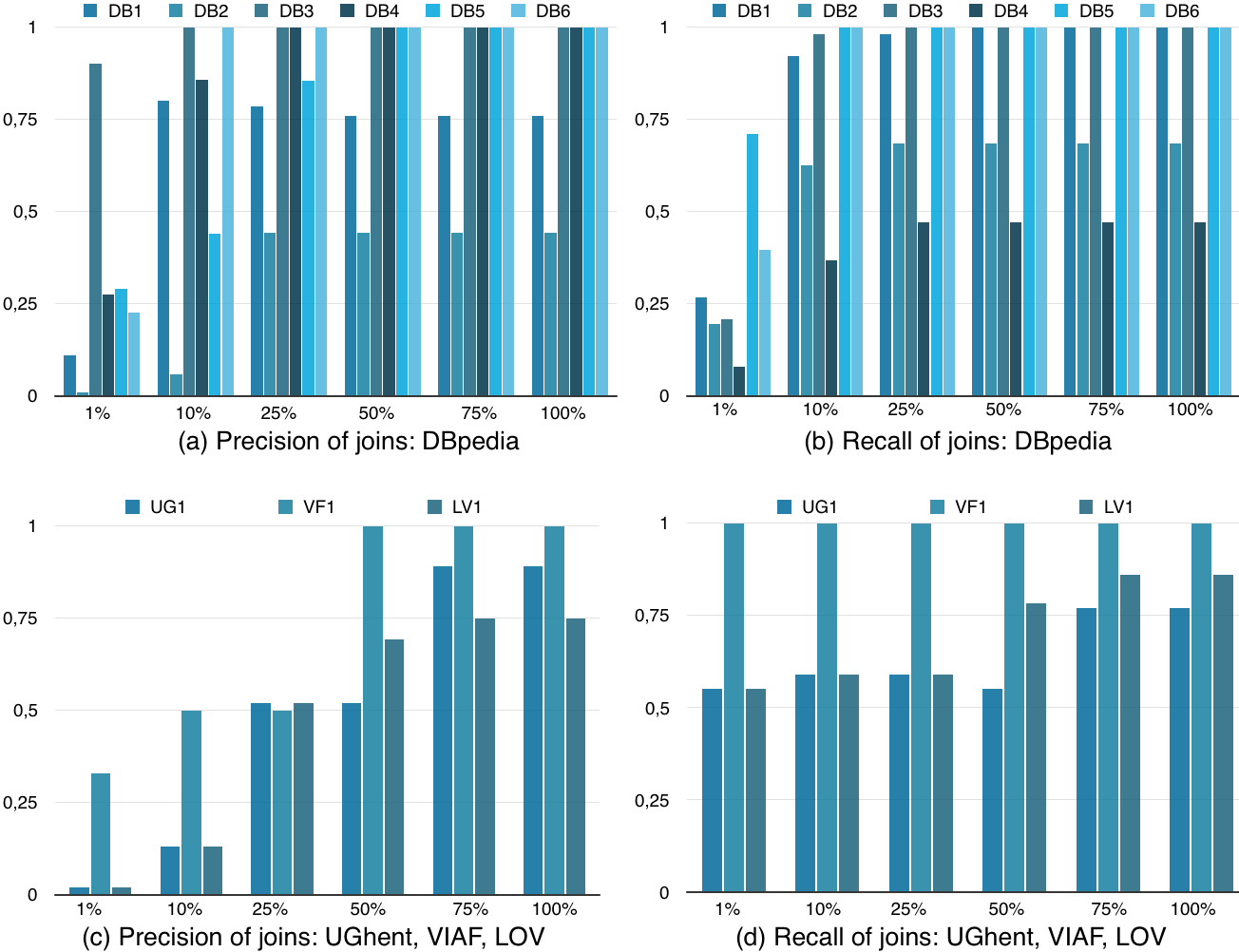}
    \caption{Precision and recall.}
    \label{fig:precisionRecall-DB-UG-VIAF-LOV}
\end{figure}

We implemented a tool to shuffle several TPF logs according to
different parameters. Thus, given $E(Q_1),...,E(Q_n)$, we are able to
produce different significant representations of
$E(Q_1\parallel...\parallel Q_n)$. 
We grouped queries targeting the same dataset into a set
of randomly chosen queries as shown in Table
\ref{tab:querysets}. For each query set, we evaluate
how $\lift(E(Q_1), gap)\cup ... \cup \lift(E(Q_n), gap) \approx
\lift(E(Q_1\parallel...\parallel Q_n), gap)$ in terms of precision and recall 
for different gap values (cf. Definition~\ref{eq:1}, Property~2). \emph{Gap} varies from 1\% to 100\% of the
log duration.
Each query set was shuffled 4 times and we calculate
the average of results by gap.
%
%
In Figure \ref{fig:precisionRecall-DB-UG-VIAF-LOV} we report results by join, (a) and (b)  show precision whereas
(c) and (d) show recall.

\rowcolors{2}{}{}
\begin{table}
\centering
  \begin{tabular}{|c|l|}
\hline
    \rowcolor{gray!70}
\textbf{Dataset} &  \textbf{Query sets}  \\  \hline
\emph{DBpedia 2015}  & 
 \begin{tabular}{ll}
$DB_1=\{${\fontsize{6}{12}\selectfont $Q_{1},~Q_{8},~Q_{14},~Q_{22}$}$\}$  & $DB_4=\{${\fontsize{6}{12}\selectfont $Q_{4},~Q_{12},~Q_{24}$}$\}$                                       \\ 
 $DB_2=\{${\fontsize{6}{12}\selectfont $Q_{3},~Q_{11},~Q_{15},~Q_{20}$}$\}$  &  $DB_5=\{${\fontsize{6}{12}\selectfont $Q_{7},~Q_{16},~Q_{21},~Q_{5}$}$\}$                                       \\ 
 $DB_3=\{${\fontsize{6}{12}\selectfont $Q_{6},~Q_{13},~Q_{19},~Q_{27}$}$\}$   & $DB_6=\{${\fontsize{6}{12}\selectfont $Q_{9},~Q_{10},~Q_{29},~Q_{30}$}$\}$                                          \\ 
 \end{tabular} \\ \hline \rowcolor{gray!25} 
 \emph{UGhent}  & $UG=\{${\fontsize{6}{12}\selectfont $Q_{2},~Q_{23},~Q_{25}, Q_{29},~Q_{30}$}$\}$  \\  \hline
\emph{LOV} & $LV=\{${\fontsize{6}{12}\selectfont $Q_{17},~Q_{18},~Q_{26},~Q_{29},~Q_{30}$}$\}$  \\  \rowcolor{gray!25} \hline
 \emph{VIAF} & $VF=\{${\fontsize{6}{12}\selectfont $Q_{29},~Q_{30}$}$\}$  \\
 \hline
\end{tabular}
\caption{Query sets executed concurrently over a TPF server.}
   \label{tab:querysets}
\end{table}

Precision and recall globally improve when gap
increases. When gap is small (less than 50\%)
precision decreases significantly. A small gap leads \lift to split values of an
inner loop i.e., the $ctpExtraction$ algorithm
cannot aggregate in one ctp all triple patterns of the inner loop.


Concerning recall,  \lift is moderately impacted by
concurrency. In some cases, \lift favours recall by producing all possible
joins in the nested-loop detection.

Concerning precision, \lift is more impacted by concurrency and
results depend on concurrent executed queries. When executed queries have triple
patterns that are semantically or syntactically similar, then \lift
generates many false joins that impact precision. 
A post-processing could filter these false joins. 


\subsection{Evaluation of \lift with real SPARQL queries}
\label{sec:endpointQueries}

The goal of this experiment is to evaluate to which extent \lift
is able to deduce BGPs of an important number of real user queries. 
We analyzed 10 hours of one day (2015-10-30) of 
the log of the DBpedia SPARQL endpoint of USEWOD 2016 dataset \cite{usewod16}.
From 380,834 http requests containing SPARQL queries, we analyzed 14,259 queries
that represent our ground truth.
We kept only SELECT queries having BGPs with more than one triple pattern 
and that return not null results. 
We made an aggregation of same queries and kept only one, by hour and by user.
That is because, if in a TPF log, the same query appears several times, \lift deduces only one query,
i.e., $\lift(E(Q_{1}) \parallel E(Q_{1}), gap) \approx BGP(Q_{1})$. 
Then, we constraint to 200 the number of queries by user for each hour. 
The OPTIONAL operator was transformed in a JOIN  and 
we filter queries having a join over predicates.
\footnote{All scripts to process logs of the DBpedia SPARQL endpoint are in Python and available at \url{https://github.com/edesmontils/BE4DBPedia/}}

We installed locally a TPF server with DBpedia 3.9 to execute and collect the TPF server logs that were
given as input to \lift. Gap was set to one hour.  
We calculate precision, recall and quality of deduced BGPs against the ground truth.
\lift returned BGPs in xml files which were loaded in a MySQL database.
This allows us to make several analysis. 

Globally, we obtained 69\% of precision, 64\% of recall and 66\% of quality.
Results are better for recurrent queries (in particular for precision) 
for which we obtained 85\% of precision, 55\% of recall and 70\% of quality.
That means that queries executed frequently under different execution contexts (i.e., concurrent queries)
are better deduced than queries requested few times.  
We analyzed results by user and calculate their averages, we obtained 84\% of precision, 
74\% of recall and  78\% of quality. From this analysis, 
we observed that there are particular users whose queries are less well
deduced, but in general \lift deductions by user are good.\footnote{These results are available at: \\ \url{http://documents.ls2n.fr/lift/dbpedia-sparql-results.tar}}


\subsection{Analysis of a real TPF log}
\label{sec:usewod}

\begin{small}
\begin{figure}
  \centering
  \begin{small}
\begin{tabular}{|l|l|} \hline
\rowcolor{gray!70} BGP[1]- deduced 126 times & BGP[2] - deduced 47 times \\ \hline
\begin{lstlisting} [
    basicstyle=\footnotesize, %\tiny or or \small or \footnotesize etc.
]
{?s_47 rdfs:label ?o_51 .
?s_47 dbpo:director ?o_52 .
?s_46 rdfs:label "Brad Pitt"@en .
?s_47 dbpo:starring ?s_46 .
?o_52 rdfs:label ?o_53 .}
\end{lstlisting}  
&
\begin{lstlisting} [
    basicstyle=\footnotesize, %\tiny or or \small or \footnotesize etc.
]
{?s_8 dbpprop:starring ?s_7 .
?s_8 rdfs:label ?o_12 .
?s_7 rdfs:label "Brad Pitt"@en .
?o_13 rdfs:label ?o_14 .
?s_8 dbpo:director ?o_13 .}
\end{lstlisting}   \\ \hline
\rowcolor{gray!70} BGP[3] - deduced 43 times & BGP[4] - deduced 34 times \\ \hline
\begin{lstlisting} [
    basicstyle=\footnotesize, %\tiny or or \small or \footnotesize etc.
]
{?s_10 rdfs:label "York"@en .
?s_11 rdf:type dbpo:Artist .
?s_11 dbpo:birthPlace ?s_10 .}
\end{lstlisting}   
&
\begin{lstlisting} [
    basicstyle=\footnotesize, %\tiny or or \small or \footnotesize etc.
]
{?s_6 dbpo:birthDate ?o_8 .
?s_6 dbpo:influencedBy dbpedia:Pablo_
                               Picasso .
?s_6 rdf:type dbpo:Artist .}
\end{lstlisting}   \\ \hline
%
\rowcolor{gray!70} BGP[5] - deduced 34 times & BGP[6] - deduced 20 times \\ \hline
\begin{lstlisting} [
    basicstyle=\footnotesize, %\tiny or or \small or \footnotesize etc.
]
{?s_9 dbpprop:cityServed dbpedia:Italy .
?s_9 rdf:type dbpo:Airport .}
\end{lstlisting}  
&
\begin{lstlisting} [
    basicstyle=\footnotesize, %\tiny or or \small or \footnotesize etc.
]
{dbpo:Agent rdfs:subClassOf ?o_13 .
?o_13 rdfs:subClassOf ?o_14 .}
\end{lstlisting}   \\ \hline
\rowcolor{gray!70} BGP[7] - deduced 17 times & BGP[8] - deduced 16 times \\ \hline
\begin{lstlisting}  
{dbpo:Activity rdfs:subClassOf ?o_29 .
?o_29 rdfs:subClassOf ?o_30 .}
\end{lstlisting} 
&
\begin{lstlisting} [
    basicstyle=\footnotesize, %\tiny or or \small or \footnotesize etc.
]
{?s_283 rdf:type dbpo:Writer .
?s_282 rdfs:label "Trinity College, 
                      Dublin"@en .
?s_283 dbpo:almaMater ?s_282 .}
\end{lstlisting}   \\ \hline
\rowcolor{gray!70} BGP[9] - deduced 15 times & BGP[10] - deduced 13 times \\ \hline
\begin{lstlisting} [
    basicstyle=\small, %\tiny or or \small or \footnotesize etc.
]
{?s_17 rdf:type dbpo:Book .
?s_17 dbpo:author ?o_18 .}
\end{lstlisting}  
&
\begin{lstlisting} [
    basicstyle=\footnotesize, %\tiny or or \small or \footnotesize etc.
]
{?s_20 dbpo:team ?o_21 .
?s_20 dbpo:birthPlace dbpedia:
        Urbel_del_Castillo .}
\end{lstlisting}   \\ \hline
\rowcolor{gray!70} BGP[11] - deduced 11 times & BGP[12] - deduced 11 times \\ \hline
\begin{lstlisting} [
    basicstyle=\footnotesize, %\tiny or or \small or \footnotesize etc.
]
{?s_33 dbpo:ingredient ?o_33 .
?s_33 dbpo:kingdom dbpedia:Plant .}
\end{lstlisting} 
&
\begin{lstlisting} [
    basicstyle=\footnotesize, %\tiny or or \small or \footnotesize etc.
]
{?s_20 rdf:type yago:Carpenter .
?s_20 rdf:type yago:PeopleExecuted
                    ByCrucifixion .}
\end{lstlisting}   \\ \hline
\rowcolor{gray!70} BGP[13] - deduced 10 times & BGP[14] - deduced 10 times \\ \hline
\begin{lstlisting} [
    basicstyle=\footnotesize, %\tiny or or \small or \footnotesize etc.
]
{?s_2 foaf:isPrimaryTopicOf ?o_3 .
?s_2 rdf:type foaf:Person .}
\end{lstlisting}  
&
\begin{lstlisting} [
    basicstyle=\footnotesize, %\tiny or or \small or \footnotesize etc.
]
{?s_35 dbpo:ingredient ?o_36 .
?s_35 dbpo:type dbpedia:Dessert .
?o_36 dbpo:kingdom dbpedia:Plant .}
\end{lstlisting}   \\ \hline
\rowcolor{gray!70} BGP[15] - deduced 9 times & BGP[16] - deduced 8 times \\ \hline
\begin{lstlisting} [
    basicstyle=\footnotesize, %\tiny or or \small or \footnotesize etc.
]
{?s_18 dbpo:team ?s_15 .
?s_15 dbpo:ground dbpedia:Urbel_del_Castillo .
?s_18 rdfs:label ?o_22 .
?s_15 rdf:type dbpo:SoccerClub .
?s_15 rdfs:label ?o_17 .}
\end{lstlisting} 
&
\begin{lstlisting} [
    basicstyle=\footnotesize, %\tiny or or \small or \footnotesize etc.
]
{?s_23 dbpprop:starring dbpedia:Brad
                               _Pitt .
?s_23 rdfs:label ?o_24 .
?s_23 dbpo:director ?o_25 .
?o_25 rdfs:label ?o_26 .}
\end{lstlisting}   \\ \hline

\rowcolor{gray!70} BGP[17] - deduced 8 times & BGP[18] - deduced 8 times \\ \hline
\begin{lstlisting} [
    basicstyle=\footnotesize, %\tiny or or \small or \footnotesize etc.
]
{?s_18 dbpo:developer ?s_16 .
?s_15 rdfs:label "Belgium"@en .
?s_16 dbpo:locationCountry ?s_15 .}
\end{lstlisting} 
&
\begin{lstlisting} [
    basicstyle=\footnotesize, %\tiny or or \small or \footnotesize etc.
]
{dbpedia:Raspberry_Pi dbpo:operating
                        System ?o_16 .
?s_17 dbpo:operatingSystem ?o_16 .
?s_17 rdf:type dbpo:Device .}
\end{lstlisting}   \\ \hline

\rowcolor{gray!70} BGP[19] - deduced 7 times & BGP[20] - deduced 7 times \\ \hline
\begin{lstlisting} [
    basicstyle=\footnotesize, %\tiny or or \small or \footnotesize etc.
]
{?s_6 rdfs:label ?o_7 .
?s_6 dbpprop:starring dbpedia:Natalie_Portman .}
\end{lstlisting} 
&
\begin{lstlisting} [
    basicstyle=\footnotesize, %\tiny or or \small or \footnotesize etc.
]
{dbpedia:Jesus dc:subject ?o_28 .
?s_29 dc:subject ?o_28 .}
\end{lstlisting}   \\ \hline

\end{tabular}
\end{small}
\caption{Recurrent BGPs extracted from the TPF log of USEWOD 2016.}
\label{fig:usewodQueries}
\end{figure}
\end{small}

The goal of this experiment is to extract BGPs from 
the log of the DBpedia's LDF server of
USEWOD 2016 dataset \cite{usewod16}. This log contains http requests
from October 2014 to November 2015. We analyzed the first quarter of
the log representing 4,720,874 single triple pattern queries (until
27th February 2015). We pruned 1\% of the log corresponding to entries that 
are not well formed TPF requests.  
\lift needs not only received triple pattern queries but also corresponding answers.
To obtain answers, we
executed the triple pattern queries of the log using a TPF client
\footnote{\url{https://github.com/LinkedDataFragments/Client.js}}.
Then, we run \lift with log slices of one hour with a maximum gap (one
hour).

We obtained 1236 BGPs containing more than one triple pattern.
Table~\ref{fig:usewodQueries}
describes the most recurrent queries. Unsurprisingly, some of them
are the queries available on the TPF web
site. 
BGP[1]
corresponds to $Q_1$, BGP[2] is like BGP[1] except that
\emph{dbpedia-owl:starring} is replaced by
\emph{dbpprop:starring}. BGP[3] corresponds to the query used as the
motivation example of~\cite{ldf_amf_metadata_15}, BGP[4]
corresponds to $Q_3$, BGP[5] to $Q_{6}$, etc.
In this top 20 list, only few queries were unknown: BGP[6], BGP[7], BGP[8], 
BGP[12] and BGP[13].
We can analyze deduced BGPs to obtain some statistics, for instance,
the most common type of join is subject-subject (2693), 
followed by subject-object (1063), what is coherent with analysis shown in \cite{gallego2011empirical}.
\lift's deductions from the TPF log of USEWOD 2016 are available 
in an xml document\footnote{\url{http://documents.ls2n.fr/lift/dbpedia-tpf-results.xml}}.



\section{Related Work}
\label{sec:related_work}

Compared to the state of art, \cite{verborgh_usewod_2015} reports statistics
on TPF logs that allow to verify server availability, number of
evaluated requests and cache usage. These statistics are useful
but can be obtained at triple pattern granularity. \lift, is 
able to extract BGPs from TPF logs, that can be subsequently analyzed to retrieve other information, 
e.g., frequently executed BGPs as done in Sections \ref{sec:endpointQueries} and \ref{sec:usewod}.

Extracting information from logs is traditionally a data mining
process~\cite{han2011data_mining}. Sequential pattern mining
\cite{mooney2013EdA-sequential} focuses on discovering frequent
subsequences from an ordered sequence of events. However, searching
for BGPs cannot be reduced to searching for frequent episodes in a TPF
log.
Suppose, two queries $Q_1:\ \{?x \ p1 \ o1 \ . \ ?x \ p2 \ ?y\}$ and
$Q_2:\ \{?x \ p1 \ ?y \ .\ ?y \ p3 \ ?z\}$. The TPF query engine
executes the joins with a nested-loop. So, $?x \ p1 \ o1$ and
$?x \ p1\ ?y$ will appear once in the log, while patterns with
$IRIs \ p2 \ ?y$ and $IRIs \ p3 \ ?z$ will appear many times according
to the selectivity of the triple patterns on $p1$. Searching for
frequent episodes will raise up episodes with $p2$ and $p3$ but joins
were between $p1, p2$ and $p1, p3$. 
Clearly, TPF logs need to be
pre-processed before analysis.

In previous
work~\cite{nassopoulos2016feta}, we focused on analyzing SPARQL logs. 
We proposed \feta, an algorithm to
reverse BGPs of federated SPARQL queries from logs collected from a
set of SPARQL endpoints. \lift and \feta share the problem statement,
but have several major differences. First, \feta analyzes a SPARQL
log gathered from many data providers in order to infer BGPs of
federated queries. \lift analyzes a single TPF log of a single data
provider to infer BGPs of SPARQL queries. This makes possible to run
\lift on real available logs as those published by USEWOD
2016 \cite{usewod16}. 
Second, the input log of \feta is composed of SPARQL queries that
are more complex than TPF queries. On the other hand,
SPARQL queries include variable names that can be used to
disambiguate some situations. In TPF logs, such variables are not
available. Next, TPFs contain metadata that helps clients to decide
a join ordering. To obtain the cardinality of a triple pattern,
a client actually requests the triple pattern making indistinguishable
these requests from regular query processing. 
This makes BGP extraction from TPF logs more challenging.


\section{Conclusions and future work}
\label{sec:conclusion}


\lift extracts BGPs from TPF server logs. 
It tracks joins executed with nested-loops following a strategy
of merging and splitting compatible triple pattern queries.
\lift deductions have good precision and good recall mainly 
when evaluating frequent queries.
The main challenge for \lift is the concurrent execution of queries 
from the same client targeting exactly the same triples (queries sharing IRIs).

\lift opens several perspectives. First, \lift focused on nested-loops. 
Clients can also rely on symmetric hash to perform
joins that can also be detected. Detecting nested-loop and symmetric hash
together from the server-side is challenging.
Second, in \lift, we favoured recall, mainly by splitting input mappings
for every intersection with output mappings. We can introduce more
constraints for splitting that should improve precision against recall, 
e.g., to separate repeated executions of identical queries 
which traces appear in the same TPF. 
Third, \lift is currently an algorithm that post-processes logs. We can
transform \lift into a streaming algorithm able to extract BGPs in
real-time. This will also allow to process very large TPF logs.
Finally, \lift is able to process logs on different datasets by just
merging them. In this case, one data provider is able to detect
cross-join against different datasets hosted on the same server. If
different data providers are ready to collaborate by just sharing
their logs, they should be able to infer federated queries running
across multiple servers.

\begin{footnotesize}
\bibliographystyle{abbrv}
\bibliography{myBib} 
\end{footnotesize}
\end{document}